\documentclass[osajnl,twocolumn,showpacs,superscriptaddress,11pt]{revtex4-1} 
\usepackage{amsmath,amssymb,graphicx,longtable,psfrag,subfigure,multirow}
\begin{document}

\title{Geometrical parameter analysis of the high sensitivity fiber optic angular displacement sensor}

\author{Jo{\~a}o M. S. Sakamoto}\email{Corresponding author: sakamoto@ieav.cta.br}
\affiliation{Division of Photonics, Instituto de Estudos Avan{\c c}ados, Trevo Cel. Av. Jos{\'e} A. A. do Amarante n$^o$ 1, S{\~a}o Jos{\'e} dos Campos, SP 12228-001, Brazil}

\author{Gefeson M. Pacheco}
\affiliation{Department of Microwave and Optoelectronics, Instituto Tecnol{\'o}gico de Aeron{\'a}utica, Pra{\c c}a Mal. Eduardo Gomes n$^o$ 50, S{\~a}o Jos{\'e} dos Campos, SP 12228-900, Brazil}

\author{Cl{\'a}udio Kitano}
\affiliation{Department of Electric Engineering, Universidade Estadual Paulista, Campus III, Ilha Solteira, SP 15385-000, Brazil}

\author{Bernhard R. Tittmann}
\affiliation{Department of Engineering Sciences and Mechanics, Pennsylvania State University, 212 Earth \& Engr. Sciences Building, University Park, PA 16802-6812, USA}

\begin{abstract}
In this work, we present an analysis of the influence of the geometrical parameters on the sensitivity and linear range of the fiber optic angular displacement sensor, through computational simulations and experiments. The geometrical parameters analyzed were the lens focal length, the gap between fibers, the fibers cladding radii, the emitting fiber critical angle (or, equivalently, the emitting fiber numerical aperture), and the standoff distance (distance between the lens and the reflective surface). Besides, we analyzed the sensor sensitivity regarding any spurious linear displacement. The simulation and experimental results showed that the parameters which play the most important roles are the emitting fiber core radius, the lens focal length, and the light coupling efficiency, while the remaining parameters have little influence on sensor characteristics.
\end{abstract}

\maketitle 

\section{Introduction}
Fiber optic angular displacement sensors have been designed for applications that require the measurement of very small angular displacements (in the order of 1 $\mu$rad or less). By measuring the angular displacement, other physical quantities can be indirectly measured or detected, as pressure, velocity, acceleration, linear displacement, and ultrasound for example~\cite{wang1997,sakamoto2012}. Therefore, this type of sensor can be used as the optical detector of an atomic force microscope (AFM)~\cite{wu1995} or for measurement of angular displacement in automotive, robotics, optics, aeronautics and aerospace industries~\cite{khiat2010}. 

\begin{figure}[ht]
  \centering
  \includegraphics[width=8.2 cm]{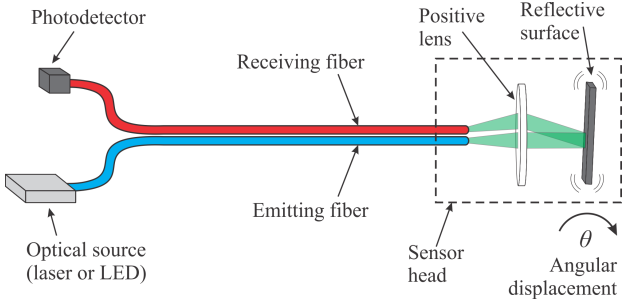}	
  \caption{Fiber optic angular displacement sensor.}
  \label{fig:SC}
\end{figure}

In recent developments~\cite{sakamoto2012,sakamoto2013}, we presented a fiber optic sensor capable of measurement of angular displacement in the order of microradians. Due to its high sensitivity, it was applied for detection of both bulk and surface ultrasonic waves.
This sensor presented very low cost and ease of assembly since it comprises only an optical source (laser or LED), two optical fibers (one emitting and the other receiving), a positive lens, a reflective surface, and a photodetector (coupled to a transimpedance amplifier), arranged as shown in Fig.~\ref{fig:SC}. 
The mathematical model (theory) of the aforementioned sensor was also depicted and validated by the comparison of the sensor's characteristic curve obtained by computational simulation and by experiment. In that case, we regarded nine different sensor configurations by varying both emitting and receiving fibers core radii~\cite{sakamoto2012}. 

In this work, using the same mathematical model, we thus performed an analysis of the influence of the geometrical parameters on the sensitivity and linear range of the fiber optic angular displacement sensor. The parameters analyzed were:
the lens focal length, the gap between fibers (or, equivalently, the fibers cladding radii), the emitting fiber critical angle (or, equivalently, the emitting fiber numerical aperture), and the standoff distance.
Besides, we analyzed the sensor sensitivity regarding any spurious linear displacement.

\section{Theory}
\begin{figure}[!ht]
  \centering
  \includegraphics[width=8.2 cm]{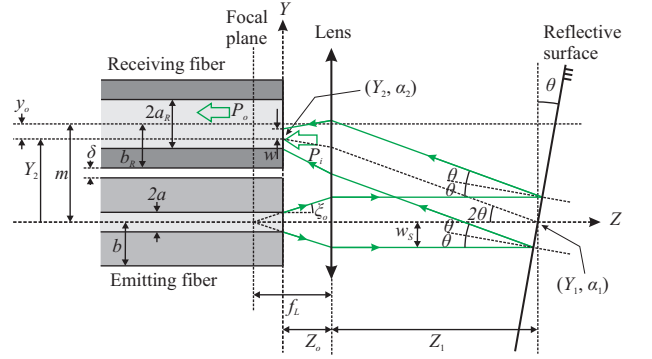}
  \caption{Sensor head detail and mathematical model variables.}
  \label{fig:SH}
\end{figure}

The mathematical model of the sensor is briefly described in this section, in order to provide the basis for computational simulations performed in this work. The sensor head components and the mathematical model variables are shown in~Fig.~\ref{fig:SH}, where:
\\$\theta$ is the angular displacement (or the reflective surface angle);\\
$a$ is the emitting fiber core radius;\\
$a_R$ is the receiving fiber core radius;\\
$b$ is the emitting fiber cladding radius;\\
$b_R$ is the receiving fiber cladding radius;\\
$N\!\!A$ is the numerical aperture of the emitting fiber;\\
$P_i$ is the total optical power incident at the fibers end plane;\\
$P_o$ is the optical power which is coupled into the receiving fiber core;\\
$y_o$ is the position of the receiving fiber center in relation to the optical spot center;\\
$w$ is the optical spot radius at the receiving fiber plane;\\
$Z_1$ is the distance between the lens and the reflective surface;\\
$f_L$ is the lens focal distance;\\
$\delta$ is the gap separation between the two fibers;\\
$n$ is the refractive index of the surrounding medium;\\
$w_s$ is the optical spot radius in the region between the lens and the reflective surface; and\\
$\xi_o$ is the critical angle.\\

The principle of operation of the sensor is based on the modulation of the intensity of light by the angular displacement (reflective surface angle), $\theta$. The power transfer coefficient, defined as the ratio between the optical output power and input power [$\eta(\theta) = P_o/P_i$], was derived as a function of $\theta$ in a previous work \cite{sakamoto2012}, and is given by:
\begin{multline}
\eta(\theta) = \frac{P_o}{P_i} = \Gamma \frac{4}{\pi w^2} \int_{0}^{y_o(\theta)+a_R}{r \exp\left(\frac{-2 r^2}{w^2}\right)} \\ \times {\cos^{-1}\left(\frac{r^2 + y_o^2(\theta) - a_R^2}{2ry_o(\theta)}\right)}{\rm d}r,
\label{eq:eta}
\end{multline}
where $r$ is the radial coordinate in relation to the optical spot center, and $\Gamma$ accounts for light coupling efficiency on the receiving fiber. The position of the receiving fiber center in relation to the optical spot center ($y_o$) is given by: 
\begin{equation}
y_o(\theta) = m - \theta 2(Z_o+Z_1-Z_1Z_o/f_L),
\label{eq:yo}
\end{equation}
where the distance between the centers of the fibers ($m$), constant, is given by:
\begin{equation}
m = b + b_R + \delta.
\label{eq:m}
\end{equation}
The distance between the lens and the fibers ($Z_o$) is set in order to collimate the light beam from the emitting fiber. Thus, it is given by:
\begin{equation}
Z_o = f_L - a/{\tan \xi_o},
\label{eq:zo}
\end{equation}
where the critical angle ($\xi_o$) is given by:
\begin{equation}
\xi_o = \sin^{-1} ({N\!\!A}/{n}).
\label{eq:xio}
\end{equation}

The optical spot radius in the region between the lens and the reflective surface ($w_s$) is given by:
\begin{equation}
\centering
w_s = f_L \tan \xi_o.
\label{eq:ws}
\end{equation}

For the particular case where $y_o=0$, the optical spot is centered on the receiving fiber core and Eq.~\ref{eq:yo} yields:
\begin{equation}
\theta=\theta_o=m/[2(Z_o+Z_1-Z_1Z_o/f_L)]. 
\label{eq:theta0}
\end{equation}

\begin{figure}[!ht]
  \centering
  \includegraphics[width=6 cm]{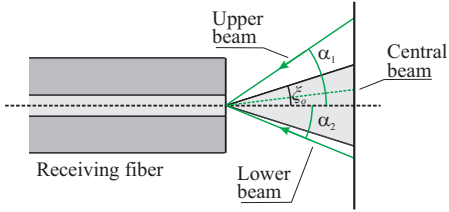}
  \caption{Detail of the receiving fiber coupling efficiency.}
  \label{fig:gamma}
\end{figure}

The light coupling efficiency $\Gamma$ on Eq.~\ref{eq:eta} is a ratio between the power accepted by receiving fiber (due to its numerical aperture, $N\!\!A_r$) and the power from the optical spot. According to Fig.~\ref{fig:gamma}, the parameter $\Gamma$ can be stated as 
\begin{equation}
\Gamma \approx \frac{2N\!\!A_r^2}{\alpha_1^2+ \alpha_2^2}, 
\label{eq:gamma}
\end{equation}
where $\alpha_1$ is the entrance angle for the upper beam on the optical spot and $\alpha_2$ is the entrance angle for the lower beam.
If the optical spot entrance angles lies inside the receiving fiber numerical aperture the coupling efficiency is maximum, while for the case where the entrance angles are outside $N\!\!A_r$, no light is coupled to the receiving fiber ($0 \leq \Gamma \leq 1$). 

The static characteristic curve for the sensor is therefore obtained by evaluating the integral on Eq.~\ref{eq:eta}, numerically, for each corresponding value of~$\theta$.

\section{Simulations and experiments}
\label{sec:simexp}
The emitting and receiving fibers core radii ($a$ and $a_R$) influences on the sensitivity and linear range were analyzed in a previous work~\cite{sakamoto2012} and are summarized in subsection~\ref{sec:coreradii}. In the present work, the influence of additional geometrical parameters as the lens focal length ($f_L$), the emitting fiber critical angle ($\xi_o$) [or, equivalently, the emitting fiber numerical aperture ($N\!\!A$)], the standoff distance ($Z_1$), the gap between fibers ($\delta$) [or, equivalently, the fibers cladding radii ($b$ and $b_R$)] were analyzed as shown in the subsections~\ref{sec:lens} to \ref{sec:Z1}.

\begin{figure}[!ht]
  \centering
  \includegraphics[width=8.2 cm]{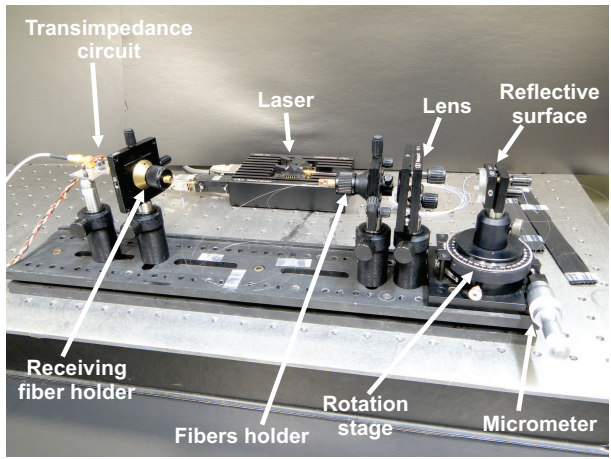}
  \caption{Fiber optic angular displacement sensor mounted in the laboratory.}
  \label{fig:sensor3rd}
\end{figure}

An experimental setup was mounted in order to acquire the static characteristic curves of the sensors with different geometrical parameters. This setup comprised an optical source, two optical fibers (one emitting and the other receiving), a positive lens, a reflective surface, and a photodetector, as shown in schematic of Fig.~\ref{fig:SC}. The laser output was delivered by an optical fiber which was spliced to the sensor's emitting fiber. The emitting fiber, in turn, was fixed parallel with the receiving fiber and their tips were aligned. The light from the emitting fiber was delivered to the lens, which collimated the beam. Thus, the collimated light impinged on the reflective surface, being modulated by the surface's angle ($\theta$) and reflected back to the lens. The lens focused the reflected beam on the receiving fiber core, which collected the light. Finally, the receiving fiber delivered the intensity modulated light to the photodetector, as shown in Fig.~\ref{fig:sensor3rd}. The photodetector was coupled to a transimpedance circuit which, in turn, was coupled to an oscilloscope. 
The reflective surface was mounted on a rotation stage with an electric micrometer (Fig.~\ref{fig:sensor3rd} shows a manual micrometer that was replaced by the electric on the experiments). This stage was controlled by a drive that allowed a continuous rotation movement, with constant velocity. Therefore, the static characteristic curve was obtained by rotating the reflective surface and acquiring the output signal on the oscilloscope.

In this work, the simulations and experiments were accomplished with a CW laser ($\lambda = 980$~nm, 68~mW) as the optical source, and a Thorlabs 980HP single mode optical fiber ($N\!\!A=0.2$, $a=a_R=1.8$~$\mu$m, $b=b_R=62.5$~$\mu$m) for both emitting and receiving fibers. Exception is Section~\ref{sec:coreradii} where additional core and cladding radii values were used on simulations.

\subsection{Optical fiber core radii}
\label{sec:coreradii}
On reference~\cite{sakamoto2012}, the static characteristic curves were simulated for nine different values of emitting/receiving core radii ($a/a_R$ in $\mu$m)
: 4/4, 4/25, 4/52.5, 25/4, 25/25, 25/52.5, 52.5/4, 52.5/25, 52.5/52.5, showing that the normalized sensitivity of the sensor increases with a decrease on the fibers core radii (for both emitting and receiving fibers) and the normalized sensitivity is dominated by the emitting fiber core radius, $a$, while the receiving fiber core radius, $a_R$, plays a secondary role. These simulations were corroborated by experiment~\cite{sakamoto2012}, showing that the mathematical model is suitable to describe the sensor behavior. Therefore, the variation on $a$ and $a_R$ in this work were only simulated using the same model. The values of $a/a_R$ on simulation were the nine combinations using fibers with 1.8~$\mu$m, 25~$\mu$m, and 52.5~$\mu$m core radii. The further parameters used in this simulation were: $f_L=6$ mm, $\delta=0$ $\mu$m, $b=b_R=62.5$ $\mu$m, $w=a+2$ $\mu$m, $Z_1=42.2$ mm, $N\!\!A=0.20$ (for $a=1.8$ $\mu$m), $N\!\!A=0.22$ (for $a=25$ $\mu$m and $a=52.5$ $\mu$m). The coupling efficiency was evaluated and used as $\Gamma=1$. The simulation results regarding the normalized sensitivity (NS) and linear range (LR) are summarized in Table~\ref{tab:core}. 
\begingroup
\squeezetable
\begin{table}[ht]
{\bf \caption{\label{tab:core}Normalized sensitivity and linear range for different emitting and receiving fibers core radii~\cite{sakamoto2012}$^{a,b}$}}
\begin{center}
\begin{tabular}{lllccc}
\hline	
\multicolumn{6}{c}{$a$ \footnotesize{[$\mu$m]}} \\
	& & & $1.8$ & $25$ & $52.5$ \\
\hline
\multirow{6}*{$a_R$ \footnotesize{[$\mu$m]}}
 & $1.8$    & NS & $(3.4\times V_{\rm max})$ & $(0.6\times V_{\rm max})$ & $(0.3\times V_{\rm max})$ \\ 
 &        & LR & 140 $\mu$rad               & 785 $\mu$rad               & 1411 $\mu$rad              \\ \cline{2-6}
 & $25$   & NS & $(2.4\times V_{\rm max})$ & $(0.4\times V_{\rm max})$ & $(0.3\times V_{\rm max})$ \\ 
 &        & LR & 176 $\mu$rad               & 1123 $\mu$rad               & 1584 $\mu$rad              \\ \cline{2-6}
 & $52.5$ & NS & $(2.5\times V_{\rm max})$ & $(0.4\times V_{\rm max})$ & $(0.2\times V_{\rm max})$ \\ 
 &        & LR & 169 $\mu$rad               & 1138 $\mu$rad               & 2063 $\mu$rad              \\
\hline
\end{tabular}
\end{center}
\footnotesize $^a$$V_{\rm max}$ must be in units of volts. \\
\footnotesize $^b$NS=Normalized sensitivity; LR=Linear range.
\end{table}
\endgroup
The NS is defined as the slope of the normalized static characteristic curve, while the LR is the maximum corresponding range on $\theta$ for which the non-linearity is less than 1\%. The variable $V_{\rm max}$ is a calibration factor that provides the sensor sensitivity (unnormalized) and it is defined as the maximum voltage (at the peak) of the static characteristic curve, i.e., the voltage measured for the angle $\theta = \theta_o$. The static characteristic curve is normalized (to provide NS) in order to compare different sensor configurations. The parameters NS and LR were numerically evaluated from the simulated static characteristic curves. 

The sensor used in the remaining part of this work was the sensor 1.8/1.8, due to its higher sensitivity. Although any other configuration instead of 1.8/1.8 could be used for these simulations without loss of generality, since they are ruled by the same equation. 

\subsection{Lens focal length}
\label{sec:lens}
Regarding the sensor configuration 1.8/1.8 (i.e., $a=1.8$ $\mu$m and $a_R=1.8$ $\mu$m), three different focal lengths were tested (6~mm, 20~mm, and 50~mm) while the other geometrical parameters were kept unchanged. The simulations for different lens focal lengths are shown in Table~\ref{tab:simfl}, where the 
operation point refers to the point where the sensor is set to detect dynamic variation of $\theta$; non-linearity refers to the maximum input deviation between the simulation curve and the linear fit used to measure the normalized sensitivity and linear range.
\begingroup
\squeezetable
\begin{table}[ht]
{\bf \caption{\label{tab:simfl}Simulation results for the lens focal length$^a$}}
\begin{center}
\begin{tabular}{p{0.4in}p{0.85in}p{0.5in}p{0.8in}p{0.4in}}
\hline	
	$f_L$ \footnotesize{[mm]}
	& Normalized sensitivity \footnotesize{[mV/$\mu$rad]}
  & Linear range \footnotesize{[$\mu$rad]}
  & Operation point \footnotesize{[V]}
  & Non-linearity \footnotesize{[\%]} \\
\hline
6 & $(3.4\times V_{\rm max})$ & 140 & $(0.6\times V_{\rm max})$ & 0.98 \\
20 & $(4.7\times V_{\rm max})$	& 94 & $(0.6\times V_{\rm max})$ & 0.95 \\
50 & $(7.8\times V_{\rm max})$	& 61 & $(0.6\times V_{\rm max})$ & 0.93 \\
\hline
\end{tabular}
\end{center}
\footnotesize $^a$$V_{\rm max}$ must be in units of volts.
\end{table}
\endgroup

Analyzing the results of Table~\ref{tab:simfl}, one can see that the normalized sensitivity increases with an increase on the lens focal length and, on the other hand, the linear range decreases. Thus, the sensor 1.8/1.8 with a focal length of 50~mm presents the highest normalized sensitivity, $(7.8~\times~V_{\rm max})$~mV/$\mu$rad, and the smallest linear range of 61~$\mu$rad. 
However, the sensitivity cannot be increased indefinitely through an increase in $f_L$, since this provides an increase in $w_s$ (Eq.~\ref{eq:ws}). For example, if the sensor is used on the detection of surface acoustic waves, the spot size $w_s$ imposes a limitation on the minimum acoustic wavelength (or maximum acoustic frequency) that can be measured. This means that a compromise between sensitivity and acoustic cutoff frequency should be taken in account. 

The experimental setup was used in order to acquire the static characteristic curves of the sensor 1.8/1.8, mounted with the following focal lengths: 6~mm, 20~mm, and 50~mm. The distance $Z_o$ was adjusted to achieve a collimated beam after the lens, according to Eq.~\ref{eq:zo}. The distance $Z_1$ was arbitrarily chosen as approximately 42~mm.

\begin{figure}[ht]
	\centering 
	\includegraphics[width=8.2 cm]{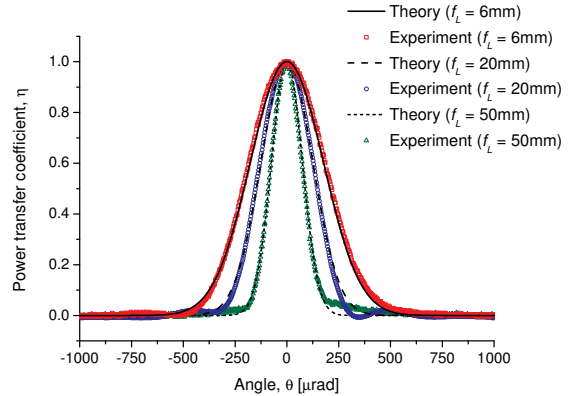}
	\caption{Simulated and experimental static characteristic curves for the sensor configuration 1.8/1.8, with: $f_L=6$~mm, $f_L=20$~mm, and $f_L=50$~mm.}
	\label{fig:grap18}
\end{figure}

The experimental results for the three sensors are shown in Fig.~\ref{fig:grap18}. In this figure, the simulation curve for $f_L=6$~mm was plotted as a solid black line and the experimental data were plotted with square red markers. Regarding $f_L=20$~mm, the simulation curve was plotted as a dashed black line and the experimental data were plotted with circle blue markers. Finally, for $f_L=50$~mm, the simulation was plotted as a short dashed black line and the experimental data were plotted with triangle green markers.

The simulation curve for $f_L =6$~mm was plotted with an optical spot of $w=a+2\mu$m, while for the sensor with $f_L=20$ mm, the optical spot radius was $w=a+8$~$\mu$m. Regarding the sensor with $f_L=50$~mm, the optical spot radius was $w=a+13$~$\mu$m. These values were adjusted to achieve a better fit with the experimental curves. Ideally, the value of $w$ should be approximately equal to the emitting fiber core radius ($w \approx a$). However, in practice this is not true due to the lens spherical aberration. The influence of spherical aberration was evaluated by computer simulation, that showed an increase in optical spot size due to the spherical aberration.
Each curve was centered on $\theta=0$ by subtracting its respective $\theta_o$ values, i.e., $\theta_o=$ 10.3~mrad, 3.1~mrad, 1.2~mrad for the sensors with $f_L=$ 6~mm, 20~mm, and 50~mm, respectively.

\subsubsection{On the influence of $\Gamma$}
\label{sec:lensgamma}
\begin{figure}[ht]
	\centering 
	\includegraphics[width=8.2 cm]{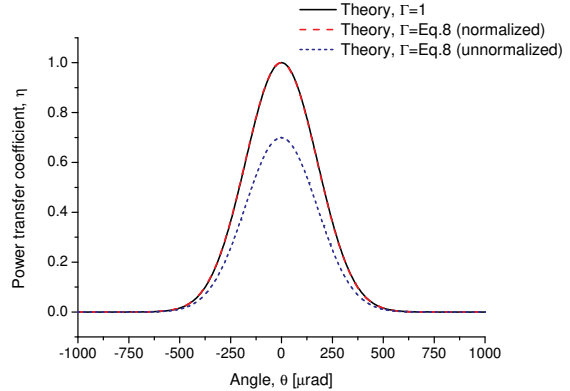}
	\caption{Simulated static characteristic curves regarding $\Gamma=1$ and $\Gamma \approx 2N\!\!A^2 /(\alpha_1^2+ \alpha_2^2)$.}
	\label{fig:lensg}
\end{figure}

The influence of the parameter $\Gamma$ was evaluated for the sensor 1.8/1.8 with $f_L=6$~mm. The graphics shown in Fig.~\ref{fig:lensg} for the theory with $\Gamma=1$ (solid black line) and the theory with $\Gamma$ as in Eq.~\ref{eq:gamma} (dashed red line), show that the shape of the curve do not change significantly (both curves are normalized). Therefore, in this case, our statement that $\Gamma=1$ is valid for the simulations. 
However, regarding the unnormalized curve with $\Gamma$ as in Eq.~\ref{eq:gamma} (short dashed blue line), it can be noted that the peak reduces around 30\%.  

\begin{figure}[ht]
	\centering 
	\includegraphics[width=8.2 cm]{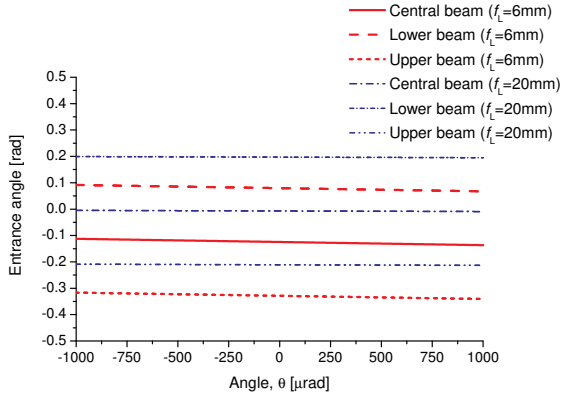}
	\caption{Entrance angles considering $\Gamma$ for sensors with $f_L=6$~mm and $f_L=20$~mm.}
	\label{fig:lensg2}
\end{figure}

In Fig.~\ref{fig:lensg2} we present a comparison between the sensors with $f_L=6$~mm and $f_L=20$~mm, regarding the entrance angle variation for the central, upper and lower beams. This figure shows that the sensor with smaller $f_L$ presents a higher variation on the entrance angles and they can lie outside the receiving fiber numerical aperture, which can reduce the peak of the characteristic curve.
Regarding an even shorter $f_L$, the peak would reduce even more (e.g. for $f_L=4$~mm, the peak presents a reduction of around 70\%), which means in practice that it would become harder to collect light on the receiving fiber, due to the coupling efficiency $\Gamma$. This imposes a practical limitation on the minimum size of the lens focal length, given a fixed $N\!\!A_r$.

\subsection{Gap between fibers and fibers cladding radii}
\label{sec:delta}
According to Eq.~\ref{eq:m}, the parameters $\delta$, $b$, and $b_R$ compose the parameter $m$, which is the distance between the centers of the fibers. In this case, the simulation was accomplished varying only $\delta$, since varying $b$ or $b_R$ would give the same effect. In addition, varying $b$ or $b_R$ would require using different fibers with different parameters, while varying $\delta$ is straightforward and keeps the other parameters unchanged. Regarding the sensor configuration 1.8/1.8 with $f_L=6$~mm, the simulation results for three $\delta$ values (0~$\mu$m, 125~$\mu$m, and 250~$\mu$m) were obtained and displayed in Table~\ref{tab:simgap}.
\begingroup
\squeezetable
\begin{table}[ht]
{\bf \caption{\label{tab:simgap}Simulation results for the gap between fibers$^a$}}
\begin{center}
\begin{tabular}{p{0.3in}p{0.85in}p{0.5in}p{0.8in}p{0.4in}}
\hline
	$\delta$ \footnotesize{[$\mu$m]}
	& Normalized sensitivity \footnotesize{[mV/$\mu$rad]}
  & Linear range \footnotesize{[$\mu$rad]}
  & Operation point \footnotesize{[V]}
  & Non-linearity \footnotesize{[\%]} \\
\hline
0 & $(3.4\times V_{\rm max})$ & 140 & $(0.6\times V_{\rm max})$ & 0.98 \\
125 & $(3.4\times V_{\rm max})$ & 140 & $(0.6\times V_{\rm max})$ & 0.98 \\
250 & $(3.4\times V_{\rm max})$ & 140 & $(0.6\times V_{\rm max})$ & 0.98 \\
\hline
\end{tabular}
\end{center}
\footnotesize $^a$$V_{\rm max}$ must be in units of volts.
\end{table}
\endgroup

\begin{figure}[ht]
	\centering 
	\subfigure[]{\includegraphics[width=4 cm]{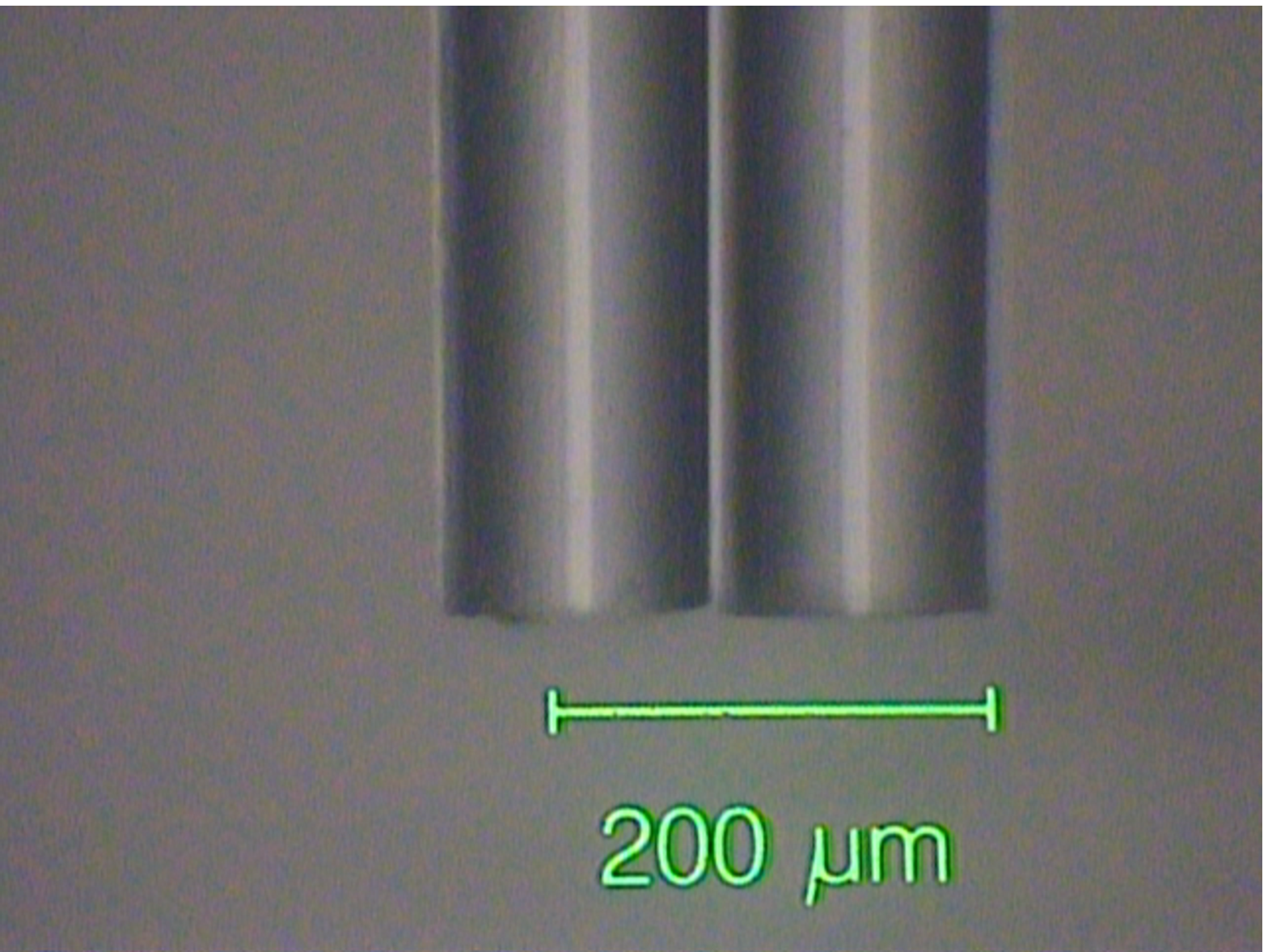}\label{fig:gap0}}\hspace{2 mm}
	\subfigure[]{\includegraphics[width=4 cm]{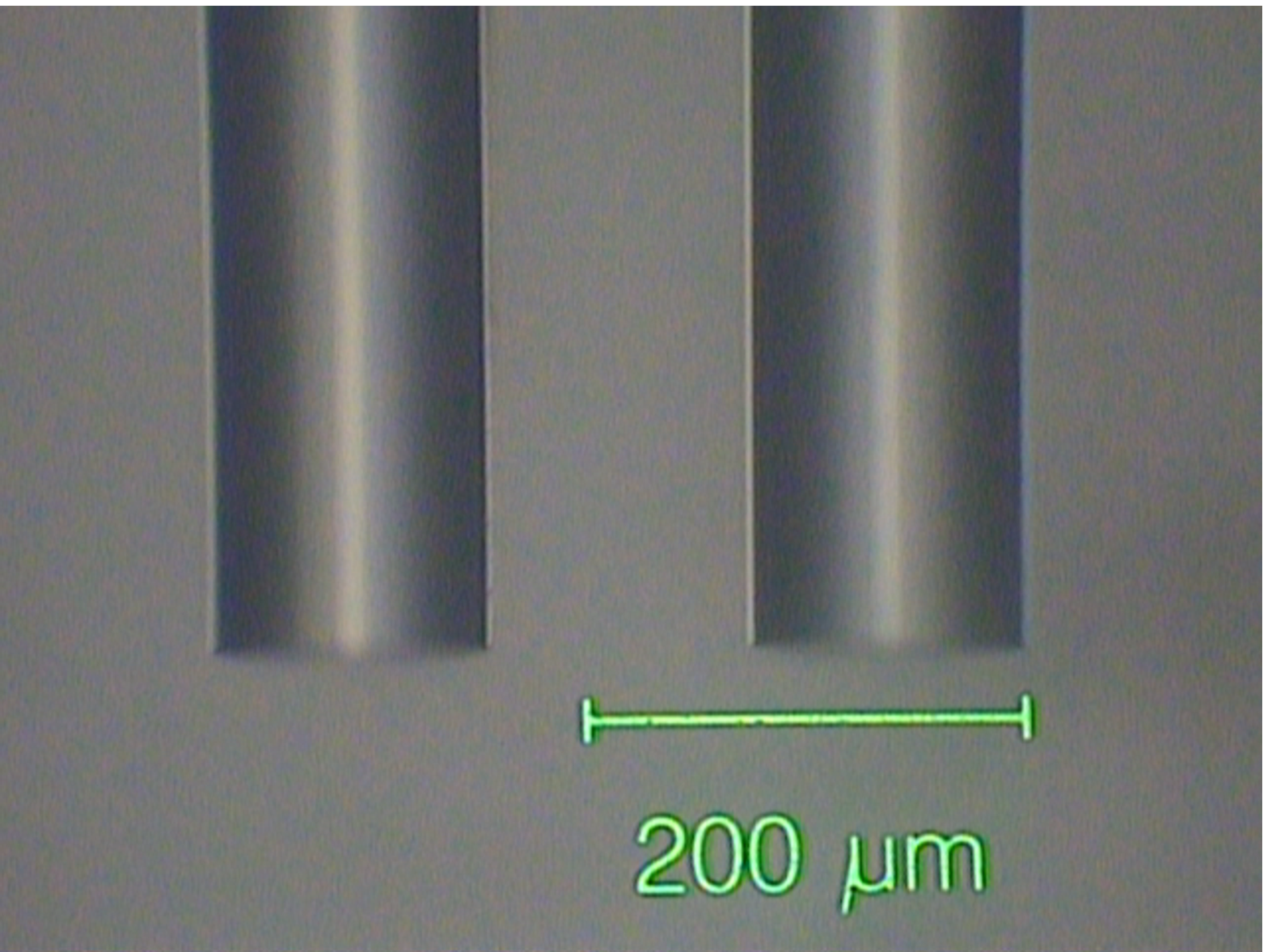}\label{fig:gap125}}
	\caption{Microscope photographs of sensors' gap. (a) Sensor without gap ($\delta = 0$ $\mu$m). (b) Sensor with gap ($\delta = 125$ $\mu$m).}
	\label{fig:photo}
\end{figure}

\begin{figure}[ht]
	\centering 
	\includegraphics[width=8 cm]{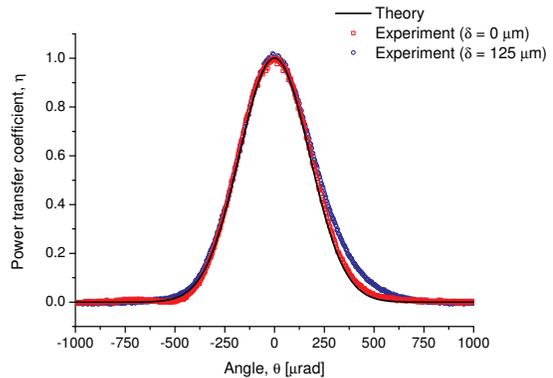}
	\caption{Static characteristic curves for the sensor configuration 1.8/1.8: theory and experiments ($\delta = 125$~$\mu$m and $\delta = 0$~$\mu$m).}
	\label{fig:delta}
\end{figure}

In order to confirm the simulation results, the sensor with gap $\delta=125$~$\mu$m was mounted with an additional idle fiber in between the emitting and receiving fibers, as shown in Fig.~\ref{fig:photo}.
The experimental characteristic curve for this sensor was acquired and it was plotted with the simulation and experiment for $\delta = 0$~$\mu$m for comparison, as shown in Fig.~\ref{fig:delta}.
The experimental result show that, in general, there is a good agreement between the two curves, which confirms the simulation results. However, it can be noted a difference on the region where 250~$\mu$rad $< \theta <$ 600~$\mu$rad, that can be attributed to the lens spherical aberration (since $m$ is larger and the light beam impinges on the lens on a region away from the paraxial one): it was observed that the mentioned region presents a broadening of the curve, which indicates that the optical spot got slight larger over this region.  

Analyzing the results in Table~\ref{tab:simgap}, one can realize that, in theory, the variation of the gap between fibers~($\delta$) and the fibers cladding radii ($b$ and $b_R$) have no influence on the normalized sensitivity or the linear range of the sensor. This is an interesting result since it shows that the mentioned parameters are not a matter of concern and the sensor head assembly could be facilitated, i.e., a small assembly error on the gap (tens of microns) would not greatly affect the sensor's performance. However, in practice, $\delta$ cannot be increased indefinitely since a large $\delta$ could prevent the proper working of the sensor or even prevent it to work, due to the finite lens diameter and lens aberrations. 

Regarding the influence of the parameter $\Gamma$ (Eq.~\ref{eq:gamma}), the peak reduction for the sensor with $\delta = 125$~$\mu$m is 53\%, while for the sensor with $\delta = 250$~$\mu$m, is 73\%. Thus in practice, increasing the value of $\delta$ also reduces the amount of light coupled on the receiving fiber.

\subsection{Numerical aperture}
\label{sec:NA}
The other geometrical parameter tested was the  emitting fiber critical angle ($\xi_o$). In this case, however, a practical value to keep in mind is its counterpart: the emitting fiber numerical aperture ($N\!\!A$). Once again, the sensor configuration 1.8/1.8 was used on the simulation where three $N\!\!A$ values were tested: 0.1, 0.2, and 0.3. The results are shown in Table~\ref{tab:simna}.
\begingroup
\squeezetable
\begin{table}[ht]
{\bf \caption{\label{tab:simna}Simulation results for the critical angle$^a$}}
\begin{center}
\begin{tabular}{p{0.3in}p{0.35in}p{0.75in}p{0.4in}p{0.8in}p{0.4in}}
\hline
	$N\!\!A$
	& $\xi_o$ \footnotesize{[rad]}		
	& Normalized sensitivity \footnotesize{[mV/$\mu$rad]}
  & Linear range \footnotesize{[$\mu$rad]}
  & Operation point \footnotesize{[V]}
  & Non-linearity \footnotesize{[\%]} \\
\hline
0.1 & 0.100 & $(2.41\times V_{\rm max})$ & 198 & $(0.58\times V_{\rm max})$ & 0.99 \\
0.2 & 0.201 & $(2.39\times V_{\rm max})$ & 194 & $(0.58\times V_{\rm max})$ & 0.96 \\
0.3 & 0.305 & $(2.38\times V_{\rm max})$ & 198 & $(0.58\times V_{\rm max})$ & 0.97 \\
\hline
\end{tabular}
\end{center}
\footnotesize $^a$$V_{\rm max}$ must be in units of volts.
\end{table}
\endgroup
According to the results in Table~\ref{tab:simna}, the variation in $N\!\!A$ has very little influence in the sensor normalized sensitivity and linear range. On the other hand, for $N\!\!A_r=0.1$ there is a peak reduction of 82\%, while for $N\!\!A_r=0.2$ the peak reduction is 30\%, and for $N\!\!A_r=0.3$, there is no peak reduction~(0\%).
Thus, although $N\!\!A$ is not a matter of concern during sensor assembly, it is better to use a higher $N\!\!A_r$, if possible.

Commercial fibers with different values for $N\!\!A$ (or $N\!\!A_r$) usually also present different values for the remaining parameters (e.g., core radius) and, therefore, an experiment regarding this parameter was not accomplished.

\subsection{Standoff distance}
\label{sec:Z1}
The sensor's standoff distance (distance between the lens and the reflective surface) was analyzed using the sensor 1.8/1.8, with $f_L=6$~mm, $\delta=0$~$\mu$m, and $N\!\!A=0.2$. The variable associated with the standoff distance is $Z_1$. Therefore, it was acquired one characteristic curve for each one of the following values: $Z_1=$ 32~mm, 57~mm, 82~mm, 107~mm, and 132~mm and the normalized results are shown in Table~\ref{tab:z1}. It is noteworthy that the $Z_1$ value was kept fixed, while $\theta$ was varied for each curve.
\begingroup
\squeezetable
\begin{table}[ht]
{\bf \caption{\label{tab:z1}Simulation results for the standoff distance$^a$}}
\begin{center}
\begin{tabular}{p{0.3in}p{0.85in}p{0.5in}p{0.8in}p{0.4in}}
\hline
	$Z_1$ \footnotesize{[mm]}
	& Normalized sensitivity \footnotesize{[mV/$\mu$rad]}
  & Linear range \footnotesize{[$\mu$rad]}
  & Operation point \footnotesize{[V]}
  & Non-linearity \footnotesize{[\%]} \\
\hline
32 & $(3.4\times V_{\rm max})$ & 140 & $(0.6\times V_{\rm max})$ & 0.98 \\
57 & $(3.4\times V_{\rm max})$ & 137 & $(0.6\times V_{\rm max})$ & 0.97 \\
82 & $(3.4\times V_{\rm max})$ & 137 & $(0.6\times V_{\rm max})$ & 0.95 \\
107 & $(3.4\times V_{\rm max})$ & 129 & $(0.6\times V_{\rm max})$ & 0.99 \\
132 & $(3.4\times V_{\rm max})$ & 126 & $(0.6\times V_{\rm max})$ & 0.99 \\
\hline
\end{tabular}
\end{center}
\footnotesize $^a$$V_{\rm max}$ must be in units of volts.
\end{table}
\endgroup

The simulation results show that the standoff distance presents little influence on sensor's normalized sensitivity and linear range. 

\begin{figure}[ht]
	\centering 
	\includegraphics[width=8 cm]{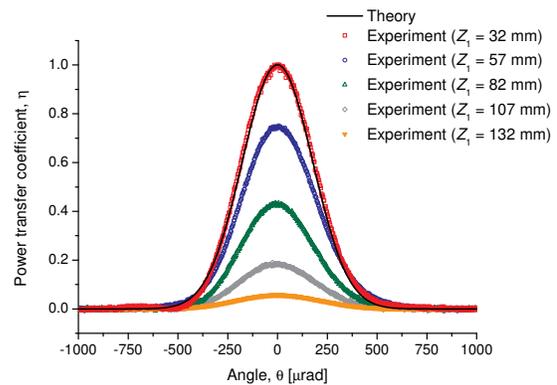}
	\caption{Static characteristic curves for the sensor configuration 1.8/1.8: theory and experiments ($Z_1 = 32$~mm, $Z_1 = 57$~mm, $Z_1 = 82$~mm, $Z_1 = 107$~mm, and $Z_1 = 132$~mm).}
	\label{fig:z1}
\end{figure}

Experimental data were acquired for these sensor configurations, shown in Fig.~\ref{fig:z1}. 
The experimental curves were normalized by the curve with $Z_1=32$~mm, instead of normalizing by itself (NS, as is the case on Table~\ref{tab:z1}). This was accomplished to point out the limitation on increasing the standoff distance: the experimental characteristic curves present a lower peak for higher $Z_1$, i.e., the sensitivity (unnormalized) decreases. This means that the sensor can lose its capability to measure $\theta$ variation, for a large standoff distance.
Differently from the statement of the previous paper~\cite{sakamoto2012} where ``Z1 can be up to approximately 750~mm'', the maximum standoff distance ($Z_1$) is smaller: around 150~mm. 

At first, the curves should keep the sensitivity. However, as will be shown in Section~\ref{sec:anglinear}, the light coupling efficiency ($\Gamma$) limits the amount of light coupled on the receiving fiber, for large standoff distances.

\subsection{Discussion}
\label{sec:dis}
The simulation results showed that the parameters that play the most important role are the emitting fiber core radius ($a$) and the lens focal length ($f_L$). The receiving fiber core radius ($a_R$) also has an influence on sensor characteristics, however, it plays a secondary role. 
The normalized sensitivity decreases with an increase on $a$ or $a_R$. On the other hand, the sensitivity increases with an increase on~$f_L$.

In order to increase the normalized sensitivity, the emitting fiber core could thus be chosen as small as possible and the receiving fiber could be both single-mode or multimode (since it plays a secondary role). In the same sense, the lens focal length could be increased. It must be observed, however, that increasing $f_L$ will also increase the spot size on the sample, $w_s$, to impractically large values, leading to a decrease on the maximum mechanical frequency that can be measured with this sensor (e.g. in a hypothetical surface acoustic wave measurement).
Another limitation in increasing the normalized sensitivity arises because the linear range decreases. Therefore, the measurement requirement will establish the limitation on minimum linear range necessary.

In addition, it was shown that the following parameters have little or no influence on the sensor's characteristics: gap between fibers ($\delta$) and emitting fiber numerical aperture ($N\!\!A$). This means that the choice of these parameters presents few restrictions. In the case of the standoff distance, in theory, there would not be restrictions on increasing it (this means that the sensor could be very far from the reflective surface). However, the experiments show that, for large $Z_1$ values, the characteristic curves present a low peak. This indicates that the unnormalized sensitivity decreases to an impractical value, limiting the maximum standoff distance, due to the light coupling efficiency.

In all cases, shown in Sections~\ref{sec:coreradii} to \ref{sec:Z1}, increasing the lens diameter would not solve the sensor's limitation (due to lens spherical aberration), since the lens distance from the fibers ($Z_o$) depends on the lens focal length ($f_L$) itself (to collimate the beam). Larger lens diameter means larger $f_L$, which means larger $Z_o$, which in turn, means larger $w_s$. A possible solution would be the use of an emitting fiber with smaller $N\!\!A$ or to setup a sensor with only one fiber to both emit and receive the light.

\section{Analysis of the sensor sensitivity for a spurious linear displacement}
\label{sec:anglinear}
The fiber optic sensor proposed in this work was designed to detect angular displacement, $\theta$. In the mathematical model we regarded that the distance between the lens and the reflective surface (or linear displacement in this case), $Z_1$, is fixed while $\theta$ is the modulation parameter. However, during a measurement of angular displacement, in a real sensor, $Z_1$ could vary, which could generate a spurious signal on the sensor output. Therefore, in order to analyze the influence of spurious variations of $Z_1$, on the power transfer coefficient, $\eta$, Eq.~\ref{eq:eta} can be rewritten as function of $Z_1$, while keeping $\theta=\theta_o$ as a constant:
\begin{multline}
\eta(Z_1) = \frac{P_o}{P_i} = \Gamma \frac{4}{\pi w^2} \int_{0}^{y_o(Z_1)+a_R}{r \exp\left(\frac{-2 r^2}{w^2}\right)} \\ \times{\cos^{-1}\left(\frac{r^2 + y_o^2(Z_1) - a_R^2}{2ry_o(Z_1)}\right)}{\rm d}r,
\label{eq:etaz1}
\end{multline}
where $y_o(Z_1)$ is given by: $y_o(Z_1)=m-2\theta_o Z_o-2\theta_o\left(1- {Z_o}/{f_L}\right)Z_1$ and $\theta_o$ is the angle which gives the maximum $\eta$, regarding an initial distance $Z_{1_o}$. This angle is given by: $\theta_o = {m}/{2(Z_o+Z_{1_o}-Z_{1_o}Z_o/f_L)}$ and the parameter $Z_{1_o}$ is defined as the starting point of the static characteristic curve. 

\begin{figure}[ht]
\centering
\includegraphics[width=8 cm]{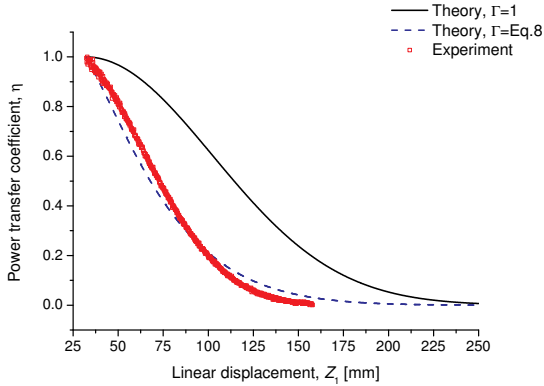}
\caption{Static characteristic curves, $\eta(Z_1)$, for the sensor 1.8/1.8. The graphic starts at~$Z_{1_o}=32$~mm.}
\label{fig:resZ1}
\end{figure}

A simulation of the static characteristic curve for the sensor 1.8/1.8, was accomplished by varying $Z_1$ on Eq.~\ref{eq:etaz1}, and keeping $\theta_o$ fixed. The parameters used were $\theta_o=10.3$~mrad and $Z_{1_o}=32$ mm. The remaining variables were kept the same as in simulations of Section~\ref{sec:simexp}, i.e., $N\!\!A = 0.2$, $a=a_R=1.8$~$\mu$m, $b = b_R = 62.5$~$\mu$m, $f_L = 6$~mm, and $w = a + 2$~$\mu$m. The simulation result (solid black line) was normalized and is shown in Fig.~\ref{fig:resZ1}, with the experimental result (square red markers).
 
The discrepancy between the theory (solid black line) and experiment (square red markers) can be attributed to the coupling efficiency ($\Gamma$, Eq.~\ref{eq:gamma}) between the receiving fiber numerical aperture ($N\!\!A_r = 0.2$) and the entrance angles from the optical spot (which change for each $Z_1$). The theoretical curve taking this parameter in account was then plotted in Fig.~\ref{fig:resZ1} as a dashed blue line, that presents a better agreement with the experiment. As can be seen, the power transfer coefficient $\eta$ vanishes for a standoff distance of approximately 150~mm, as shown by the experiments in Section~\ref{sec:Z1}.

In order to determine the sensor parameters, a linear curve was fitted to the experimental static characteristic curve and the results are shown in Table~\ref{tab:z1sp}.
\begingroup
\squeezetable
\begin{table}[ht]
{\bf \caption{\label{tab:z1sp}Results for a spurious variation on $Z_1$ $^a$}}
\begin{center}
\begin{tabular}{p{0.7in}p{0.85in}p{0.5in}p{0.8in}p{0.4in}}
\hline
	Curve
	& Normalized sensitivity \footnotesize{[mV/mm]}
  & Linear range \footnotesize{[mm]}
  & Operation point \footnotesize{[V]}
  & Non-linearity \footnotesize{[\%]} \\
\hline
Experiment & $(-13.4\times V_{\rm max})$ & 34 & $(0.6\times V_{\rm max})$ & 0.98 \\
\hline
\end{tabular}
\end{center}
\footnotesize $^a$$V_{\rm max}$ must be in units of volts.
\end{table}
\endgroup

\begin{figure}[ht]
	\centering 
	\includegraphics[width=8 cm]{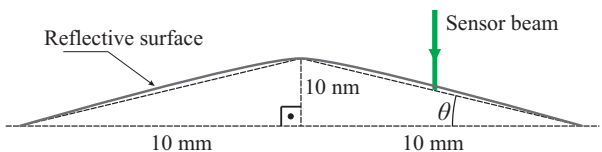}
	\caption{Angular displacement of the surface (out of scale).}
	\label{fig:tilt}
\end{figure}

To compare the parameter $Z_1$ versus the parameter $\theta$, let's consider $V_{\rm max}=1$ V. Thus, the sensitivity (unnormalized) for $Z_1$ is $\partial \eta /\partial Z_1=-13.4$~mV/mm, while for $\theta$ is $\partial \eta /\partial \theta=3.4$~mV/$\mu$rad.
Regarding a linear displacement, $Z_1$, of $10$~nm (a typical ultrasound amplitude), the output voltage would be extremely low, about $-134$~nV.
On the other hand, regarding that the same displacement, $10$~nm, tilts the surface in such way that the basis is about $20$~mm, the angle $\theta$ is then given by: $\theta \approx \tan\theta = 1$~$\mu$rad, as shown in Fig.~\ref{fig:tilt}.
Therefore, under the point of view of the angular displacement, a 1~$\mu$rad angle would provide an output voltage of 3.4~mV, i.e., the sensitivity for angular displacement is much higher (about four magnitude orders) than the sensitivity for linear displacement.

Based on this result, this sensor is not recommended for a hypothetical use to measure linear displacement on small scale (on the order of nanometers), due to its low sensitivity to linear displacement. Neither large scale (on the order of millimeters) is recommended, since a spurious angular displacement would provide a much higher output than the linear displacement itself, i.e., the output voltage would not correspond to the actual linear displacement. That is the reason why this sensor is suitable to measure angular displacement and not linear displacement.

\section{Conclusion}
The simulation and experimental results showed that the fiber optic angular displacement sensor can be constructed according to specifications of sensitivity and linear range (i.e., according to the measurement requirements) by choosing the proper geometrical parameters. Alternatively, the sensor characteristics can be calculated for a given set of geometrical parameters. 

The sensitivity and linear range suffer low influence from the gap between fibers and the numerical aperture of the emitting fiber. In contrast, the geometrical parameters that can cause large variations on sensor's characteristics are the core radii~\cite{sakamoto2012} and the lens focal distance. The light coupling efficiency on the receiving fiber limits the maximum standoff distance (pointed out by the experiments as around 150~mm) and the unnormalized sensitivity.

The sensor presented a higher sensitivity to angular displacement ($\partial \eta /\partial \theta=3.4$~mV/$\mu$rad) than the sensitivity to linear displacement ($\partial \eta /\partial Z_1=-13.4$~mV/mm), since the difference on the output voltage can be of about four magnitude orders for the same displacement. This uncoupling between linear and angular displacements gives an improvement in relation to optical sensors that are simultaneously sensitive to both~\cite{sagrario1998}, as the knife-edge sensor~\cite{monchalin1986}, the optical beam deflection sensor~\cite{murfin2000} or the fiber optic linear displacement sensor~\cite{cao2005}. 

\section*{Acknowledgments}
The authors would like to thank the Brazilian funding agencies CNPq and FAPESP (INCT de Estudos do Espa{\c c}o, 2008/57866-1) for partial funding of this research. One of the authors (JMSS) acknowledges the funding agencies CNPq (142191/2007-8) and CAPES (4697/08-1) for the provision of scholarships, and FAPESP for complementary funding (2013/07784-7). The authors recognize the contributions of Dr. Marcelo G. Destro, Dr. Kelly C. Jorge, and Dr. Rudimar Riva.

\end{document}